\documentstyle[12pt]{article}
\newcommand{\newc}{\newcommand}
\newc{\R}{$R$}
\newc{\ru}{$U(1)_R$\ }
\newc{\fun}[4]{\kz{1}^{#1}\kz{2}^{#2}\kz{3}^{#3}\kz{4}^{#4}}
\newc{\funth}[3]{\kz{1}^{#1}\kz{2}^{#2}\kz{4}^{#3}}
\newc{\funtw}[3]{\kz{1}^{#1}\kz{3}^{#2}\kz{4}^{#3}}
\newc{\funfo}[3]{\kz{1}^{#1}\kz{2}^{#2}\kz{3}^{#3}}
\newc{\funtwth}[2]{\kz{1}^{#1}\kz{3}^{#2}}
\newc{\del}{\partial}
\newc{\vev}[1]{<\!{#1}\!>}
\newc{\beq}{\begin{equation}}
\newc{\eeq}{\end{equation}}
\newc{\barr}{\begin{eqnarray}}
\newc{\earr}{\end{eqnarray}}
\newc{\ra}{\rightarrow}
\newc{\lam}{\lambda}
\newc{\eps}{\epsilon}
\newc{\half}{\frac{1}{2}}
\newc{\third}{\frac{1}{3}}
\newc{\fourth}{\frac{1}{4}}
\newc{\eighth}{\frac{1}{8}}
\newc{\gev}{\,GeV}
\newc{\lra}{\leftrightarrow}
\newc{\Dslash}{\not\!\! D}
\newc{\sg}{{\cal G}}
\newc{\eq}[1]{(\ref{eq:#1})}
\newc{\eqs}[2]{(\ref{eq:#1},\ref{eq:#2})}
\newc{\etal}{{\it et al.}\ }
\newc{\Hbar}{{\bar H}}
\newc{\hhbar}{{\overline h}}
\newc{\Ubar}{{\bar U}}
\newc{\Dbar}{{\bar D}}
\newc{\Ebar}{{\bar E}}
\newc{\eg}{{\it e.g.}\ }
\newc{\ie}{{\it i.e.}\ }
\newc{\nonum}{\nonumber}
\newc{\kap}{\kappa}
\newc{\kapi}{\frac{1}{\kap}}
\newc{\kz}[1]{(\kap z_{#1})}
\newc{\lab}[1]{\label{eq:#1}}
\newc{\oc}{{\cal O}}
\newc{\vecl}{\vec{l}}
\newc{\lle}[3]{L_{#1}L_{#2}\Ebar_{#3}}
\newc{\lqd}[3]{L_{#1}Q_{#2}\Dbar_{#3}}
\newc{\udd}[3]{\Ubar_{#1}\Dbar_{#2}\Dbar_{#3}}
\begin{document}
\vspace*{1.5ex}
{\centerline{\large{\bf Anomaly-Free Gauged R-Symmetry}}
\vspace*{6.0ex}
{\centerline{\large Herbi K. Dreiner}}
\vspace*{1.5ex}
{\centerline{\large{\it Theoretische Physik, ETH-Z\"urich, CH-8093 Z\"urich}}}
\vspace*{1.5ex}
{\centerline{\large{Work done in collaboration with A. Chamseddine}}}
\vspace*{4.5ex}
{\centerline{\bf Abstract}}

\noindent We review the gauging of an R-symmetry in local and global susy. We
then construct the first anomaly-free models. We break the R-symmetry and susy
at the Planck scale and discuss the low-energy effects. We include a solution
to the mu-problem, and the prediction of observable effects at HERA. The
models also nicely allow for GUT-scale baryogenesis and R-parity violation
without the sphaleron interactions erasing the baryon-asymmetry.

\medskip

{\bf 1.} {\bf Introduction} \newline
\R-symmetries have been widely employed as discrete and global symmetries in
susy. It is the purpose of this talk to discuss local anomaly-free
R-symmetries. This paper is similar in spirit to \cite{u1p} except we
consider \R-symmetries. It is based on the work \cite{us}.
For global susy theories the global
\R-transformations are \cite{rinvariance1}
\beq
\begin{array}{ccc}
V_k(x,\theta,{\bar\theta})& \ra&
V_k(x,\theta e^{-i\alpha},{\bar\theta}e^{i\alpha}), \\
S_i(x,\theta,{\bar\theta}) &\ra& e^{in_i\alpha}
S_i(x,\theta e^{-i\alpha},{\bar\theta}e^{i\alpha}),
\end{array}
\lab{grtrans}
\eeq
where $V_k$ is a gauge vector multiplet $S_i$ are left-handed chiral
superfields. All gauginos transform non-trivially and with the {\it same}
charge. The scalar fermions transform differently from their fermionic
superpartners.  The action for the superpotential $\int d^2\theta\,g(S_i),$ is
invariant provided
\beq
g(S_i)\ra e^{-2i\alpha}g(S_i).
\lab{rsuper}
\eeq
It is {\it not} possible in global susy theories to promote the
global \R-invariance to a local one. (a)~When the \R-parameter $\alpha$
becomes local then
\beq
\theta\ra\theta e^{-i\alpha(x)},\quad {\bar\theta}\ra{\bar\theta}
e^{i\alpha(x)},
\eeq
which is a {\it local} superspace transformation.
(b) For a local \R-symmetry the \R\ gauge vector boson $V_\mu^R$ couples to
the \R-gauginos $\lam^R$
\beq
{\cal L}\sim
{\overline \lam^R_L}(\del_\mu-ig_R V_\mu^R)\gamma^\mu\lam_L^R
+{\overline \lam^R_R}(\del_\mu+ig_R V_\mu^R)\gamma^\mu\lam_R^R.
\lab{rgaugino}
\eeq
So $g_R{\overline\lam^R}\gamma^\mu\gamma_5\lam^R V_\mu^R,$ must be in the
Lagrangian but it isn't. In order to construct a susy Lagrangian containing
this we must consider its susy transformation. It contains the term $g_R
\eps^{\mu\nu\rho\sigma}{\overline\eps}\gamma_\mu\lam^R F_{\nu\rho}^R
V_\sigma^R=\eps^{\mu\nu\rho\sigma}\delta V_\mu^RV_\nu^R F_{\rho\sigma}^R,$
since the susy variation of the gaugino term $\delta\lam^R$ contains
$\gamma^{\mu\nu}\eps F_{\mu\nu}^R$. This can not be cancelled without
departing from global susy. (c) The \R-symmetry generator \R\
does not commute with the susy generator $Q$
\beq
\left[Q_\alpha,R \right] = i(\gamma_5)_\alpha^\beta Q_\beta.
\eeq
The above equation can only hold for local \R\ if the susy algebra is local.

{\bf 2.} {\bf Local Susy}\newline
We generalize the \R-symmetry to the graviton multiplet as
\beq
e^m_\mu\ra e^m_\mu, \quad \psi_\mu\ra\exp(-i\alpha\gamma_5)\psi_\mu.
\lab{rgravity}
\eeq
The \R-gauge boson couples axially to the gravitino, the
gauginos, and the chiral fermions. Such a Lagrangian was first constructed by
Freedman \cite{freedman}. The variation of (4) is now cancelled by
\beq
e^{-1}{\cal L}=\frac{i}{\sqrt{2}}
{\overline\psi}_\rho \gamma^{\mu\nu} F_{\mu\nu}^R\gamma^\rho\lam^R,
\eeq
in the action since $\delta\psi_\mu$ contains $g_RV^R_\mu\gamma_5\epsilon$.
Ferrara \etal\cite{ferkugo} showed that any \R-invariant gauged
action can be put into the canonical form of local susy with the function
${\sg}(z_i,{\bar z}^i)= 3 \ln (\third \phi(z_i,{\bar z}^i))-\ln|g(z_i)
g^{*}(z^i)|.$ The non-invariance of $\ln|g(z_i)g^{*}(z_i)|$ under
\R\ implies the appearance of the Fayet-Illiopoulos term in the D-term
\barr
g_R\,{\sg}_{,}^{\,i}\,n_iz_i&=&g_R(3\frac{\phi^{\,i}_,}{\phi}-
\frac{g_,^{\,i}}{g})n_iz_i.\\
n_iz_ig_,^i&=&3\xi g.
\lab{fiterm}
\earr
This leads to a cosmological constant of order $\kap^4$ which fixes the
scale of \ru-breaking.

\medskip

{\bf 3.} {\bf Conditions for the Cancellation of Anomalies}\newline
\indent{\bf 3.1} {\bf Family Independent Gauged R-symmetry}\newline
We construct a anomaly-free $N=1$ local susy theory with the
gauge group $G_{SM}\times U(1)_R \equiv SU(3)_C\times SU(2)_L\times
U(1)_Y\times U(1)_R$. The matter chiral multiplets are
\barr
L:&&(1,2,-\frac{1}{2},l),\quad\Ebar: (1,1,1,e),\quad
Q:(3,2,\frac{1}{6},q), \quad \Ubar:{\bar3},1,-\frac{2}{3},u),\\
\Dbar:&& ({\bar3},1,\frac{1}{3},d),\quad H: (1,2,-\frac{1}{2},h),\quad
\Hbar: (1,{\bar2},\frac{1}{2},{\bar h}), \quad
N:  (1,1,0,n),\quad z_m:  (1,1,0,z_m),\nonumber
\lab{qmnumbers}
\earr
where we have indicated the gauge quantum numbers.
The \ru  quantum numbers are for the chiral fermions.
The superpotential in the observable sector has the form
\beq
g^{(O)}= h_{E}^{ij}L_i\Ebar_jH + h_{D}^{ij}Q_i\Dbar_jH +
h_{U}^{ij}Q_i\Ubar_j\Hbar+ h_N NH\Hbar,
\lab{superobs}
\eeq
where $h_E,\,h_D,\,h_U,\,h_N$  are the Yukawa couplings. We assume the theory
conserves \R-parity. The requirement that comes from \R-invariance for
$g^{(O)}$ is
\barr
l+e+h&=&-1, \quad q+d+h=-1,\quad q+u+{\bar h}= -1,\quad
n+h+{\bar h}=-1.\lab{muterm}
\earr
The equations for the absence of the $U(1)_Y-U(1)_R$ anomalies give
\barr
C_1\equiv 3[\frac{1}{2}l+e+\frac{1}{6}q+\frac{4}{3}u+\frac{1}{3}d] +
\frac{1}{2}(h+{\bar h}) &=&0, \lab{yyr}\\
3[-l^2+e^2+q^2-2u^2+d^2]-h^2+{\bar h}^2 &=&0, \lab{yrr}\\
3[2l^3+e^3+6q^3+3u^3+3d^3]+ 2h^3+2{\bar h}^3+16
+n^3+\sum z_m^3&=&0. \lab{rrr}
\earr
The term $16=13+3$ is due to the 13 gauginos as well as the gravitino. The
absence of the mixed $U(1)_R\,SU(2)_L$ and $U(1)_R-SU(3)_C$ anomalies implies
\barr
C_2&\equiv&3[\frac{1}{2}l+\frac{3}{2}q]+\frac{1}{2}(h+{\bar h}) +2 = 0.
\lab{sulr}\\
C_3&\equiv& 3[q+\frac{1}{2}u+\frac{1}{2}d]+3=0.
\lab{sucr}
\earr
The cancellation of the mixed gravitational anomaly \cite{gravity} requires
$Tr R=0,$
\beq
3[2l+e+6q+3u+3d]+2(h+{\bar h})-8+n+\sum z_m=0.
\lab{gravr}
\eeq
The term $-8=13-21$ is due to the 13 gauginos as well as the gravitino.
These ten equations do not have a solution, independently of the singlet
charges.

{\bf 3.2} {\bf Green-Schwarz Anomaly Cancellation}\newline
The Green-Schwarz mechanism of anomaly cancellation relies on coupling the
system to a linear multiplet $(B_{\mu\nu},\phi,\chi)$ where $B_{\mu\nu}$ is an
antisymmetric tensor \cite{gs}. The  non-invariant part of the gauge
transformations of
the action of $B_{\mu\nu}$ are of exactly the same form as the mixed gauge
anomalies $C_1,C_2,$ and $C_3$. The combined action is gauge invariant
provided $C_1/k_1=C_2/k_2=C_3/k_3.$ The $k_i$ are the Ka\v{c}-Moody
levels of the gauge algebra. For $k_2=k_3$ the anomaly cancellation
conditions are compatible if $C_2=C_1+6.$ We can simplify the equations by
assuming that ${C_2}/{C_1}={3}/{5},$ ($\sin^2\theta_w
=\frac{3}{8}$). Then $C_1= -15,\; C_2=C_3\,=-9.$
The anomaly cancellation equations can all be expressed in terms
of one variable $l'=\frac{30}{7}\cdot l$ beyond the quantum numbers of
the singlet fields $z_m$. The remaining equations (including the linear
multiplet) are
\barr
-80+\frac{3}{2}l'+\sum z_m&=&0, \quad
-\frac{8004}{9}-24l'+\frac{19}{5} {l'}^2+\frac{3}{8} {l'}^3
+\sum z_m^3 =0,
\earr
There is no rational solution for zero or one singlet.
We have performed a numerical scan for three singlets and found no solution.
We conclude that it is not possible to cancel the anomaly via the
Green-Schwarz mechanism with a small number of singlets.

{\bf 3.3} {\bf Non-Singlet Field Extensions}\newline
We allow for extra generations $N_g$ and pairs of Higgs doublets $N_h$. The
anomaly equations are
\beq
\begin{array}{rclrclrcl}
h&=&-(l+e+1),\quad &{\bar h}&=&l+e-1,\quad &q&=&-\frac{2}{9}-
\frac{1}{3}l,\\
d&=&\frac{2}{9}+\frac{4}{3}l+e,&u&=&\frac{2}{9}-\frac{2}{3} l -e,
 &n&=&1, \qquad \;\;o_c=-1.
\lab{occharg}
\end{array}
\eeq
\vspace{-0.8cm}
\barr
3(2l+e)-19+\sum_i z_i &=&0,\quad 3(2l+e)^3+13+\sum_i z_i^3=0.
\earr
We found many solutions with four singlets, \eg
$(2l+e,z_1,z_2,z_3,z_4)=(1,-\frac{47}{3},-\frac{25}{3},3,13).$
The fermionic component of the octet chiral
superfield has $R$-charge $-1$ and the scalar potential of the octet
is unconstrained and typically breaks $SU(3)_c$.

{\bf 3.4} {\bf Family Dependent Gauged \ru Symmetry}\newline
We denote the \R-quantum number of the matter fields by $e_i,l_i,q_i,u_i$,
and $d_i$, $i=1,2,3$. We assume a left-right symmetry
\beq
e_i=l_i,\quad u_i=d_i=q_i,\quad i=1,2,3.\lab{lrcharg}
\eeq
We assume that only the fields of the third generation enter the
superpotential.
\beq
g^{(O)}= h_{E}^{33}L_3\Ebar_3H + h_{D}^{33}Q_3\Dbar_3H +
h_{U}^{33}Q_3\Ubar_3\Hbar+ h_N NH\Hbar.\lab{superfamdep}
\eeq
The masses for the first and second generation will be generated
after the breaking of some symmetry, possibly the \R-symmetry. The anomaly
equations are solved in the visible fields and reduce to
\beq
\begin{array}{rcll}
h={\bar h}&=&-1,&q_3=l_3=0 ,\\
l_2&=&\frac{5}{2}-l_1,\qquad
&q_2=-(\frac{3}{2}+q_1), \qquad
n=1.
\end{array}
\lab{famdepsolns}
\eeq
\vspace{-0.5cm}
\barr
\frac{45}{2}l_1(l_1-\frac{5}{2})-54q_1(q_1+\frac{3}{2})+\frac{155}{8}+
\sum z^3_m&=&0,\quad \sum z_m=\frac{43}{2}.\lab{fdlin}
\earr
For one singlet we find two solutions. The charge of the singlet is positive
which leads to an unacceptable cosmological constant. Some of the fermionic
charges of the observable fields are $<-1$. The potential then requires
fine-tuning to guarantee weak-scale sfermion masses. For two singlets we find
many solutions. These solutions have negative singlet charges but $q_1$ or
$q_2<-1$. We found one three singlet solution.
\beq
\left\{(q_1,q_2,q_3);(l_1,l_2,l_3);(z_1,z_2,z_3)\right\}=
\left\{(-1,-\half,0);(\half,2,0);(-\frac{115}{3},
26,\frac{203}{6})\right\}. \lab{3sing}
\eeq
There are three further solutions obtained by
$q_1\lra q_2$ and $l_1\lra l_2$.
For four singlets we find many solutions. The solutions with observable
field fermionic charges greater than $-1$ are
\barr
q_1\,=\,-1,\quad l_1&=&\frac{n}{6}, \quad n=-6,...,6,\,n\not=0\lab{soludd}\\
q_1\,=\,-\frac{5}{6},\quad l_1&=&\frac{n}{6},\quad n=-6,...,6,\, n\not=-4,0,4
\lab{solgen}
\earr
The other charges are given in \cite{us}.
The solutions with $q_1=-1$ has an unacceptable level of proton decay.

{\bf 4.} {\bf Susy and R-symmetry Breaking}\newline
To have a realistic model both susy and \R-symmetry must be broken at
low energies. A Fayet-Illiopoulos term is necessarily present in the
D-term\footnote{Here we have assumed that the kinetic
energy is minimal and of the form $y=\frac{\kap^2}{2}z_iz^i+...$}
and we have a cosmological constant of the order of the Planck scale. In a
realistic model to lowest order the condition
\beq
\vev{n_iz^iz_i}+\frac{4}{\kap^2}=0,
\eeq
must be satisfied \cite{alicl2}. At least one chiral superfield must have
negative \R-charge. Only the singlets should get a vev at the Planck scale.
The most general polynomial with \R-charge 2 for the three singlet solution
is given by
\barr
g'(z_1,z_2,z_3)&=&\frac{1}{\kap^3}\left(a_1\kz{1}^{10}\kz{2}
\kz{3}^{10}+a_2 \kz{1}^{25}\kz{2}^{14}\kz{3}^{16}\right.
\\ &&\left. +a_3\kz{1}^{33}\kz{2}^{7}\kz{3}^{30}+a_4 \kz{1}^{41}
\kz{3}^{44}+ ....\right).
\earr
We take the arbitrary parameters
$a_k=\oc(1)$.
We can not break susy via the Polonyi mechanism since a constant is not
\R-invariant. We need at least three non-zero parameters $a_k$ in $g'$ then
it is possible to find solutions for which the total potential $V$ is positive
semi-definite with the value zero at the minimum, and where the D-term is also
zero at the minimum. The \R-gauge vector boson mass is then of order the
Planck mass. The total superpotential and potential are
\beq
g=g'(z_1,z_2,z_3)+g^{(O)}(S_i),
\eeq
\barr
V&=&\frac{1}{\kap^4} e^{{\sg}}\left( {\sg_,^{-1 a}}_b \sg_{,a}
{\sg_,}^b-3\right) +\half {\tilde g}^2 {\cal R}{e} f_{\alpha\beta}^{-1}
\left(\sg_,^a(T^\alpha z)_a\right) \left(\sg_,^b(T^\beta z)_b\right).
\earr
For the three-singlet model we thus obtain the D-term as
\beq
g_R^2\frac{1}{8}\left(\frac{2}{3}\right)^2\left(
-\frac{112}{3}|z_1|^2+27|z_2|^2+\frac{209}{6}|z_3|^2+\frac{4}{\kap^2}
\right)^2
\eeq
In $g'$ it is clear that there is no symmetry in $z_1,z_2,z_3$
and their vevs will be unequal. For the D-term to vanish at the minimum we
must have $|z_2|<\frac{112}{81}|z_1|$, and $|z_3|<\frac{224}{209}|z_1|$. By
fine-tuning the parameters $a_k$ it might be
possible to arrange for $|z_2|\approx z_3\approx\half|z_1|$ so that $|z_1|
\approx\frac{1}{\sqrt{5}}\kapi$. Then if we start with the natural
Planck scale $\kapi$, the effective value of $g'$ will be
$\frac{m_s}{\kap^2}$, where $m_s=\frac{1}{\kap^2}\left(\frac{1}{5}
\right)^{21}\left(\half\right)^{11}$ is of order $\oc(10^2\gev)$.
We shall assume that $z_1,z_2,z_3\approx\oc
(\kapi)$ with coefficients less than one, so that when these fields are
integrated out one gets $\vev{\kap^2 g'}=m_s$.

By integrating the hidden sector fields $z_1,z_2,z_3$ one obtains the
effective potential as a function of the light fields $z_i$. It was shown in
\cite{cfm} that the low-energy effective potential is identical to that of
the MSSM
\barr
V&=& |{\hat g}_{,i}|^2 +m_s^2|z_i|^2+m_s\left(z_i {\hat g}_{,i}
+(A-3){\hat g} +h.c.\right) \nonum \\
&&+\frac{1}{8}g^2\left( H^*\sigma^aH+\Hbar^*\sigma^a\Hbar\right)^2
+\frac{1}{8}{g'}^2\left( H^*H-\Hbar^*\Hbar\right)^2
\earr
The three singlet solution is problematic with the $\Ubar\Dbar\Dbar$ couplings
as will be clear in the next section. Therefore we must consider the four
singlet solutions which we required to avoid such a problem. The
superpotentials for the ten different classes are given in Table~2.

As before we have to tune the parameters $a_k$ so that the potential is
positive definite and so that $|z_1|,...,|z_4|\approx\oc(\kapi)$ with
coefficients less than one so as to induce a scale such that $\vev{\kap^2 g'}
=m_s=\oc(10^2\gev)$. The effective potential takes the same form as in the
three singlet case, but with different \R-numbers for the squarks and
sleptons.

{\bf 5.} {\bf Applications to R-parity Violation}\newline
When extending the Standard Model to susy new dimension four Yukawa
couplings are allowed which violate baryon- and lepton-number.
\beq
L_iL_j\Ebar_k,\quad L_iQ_j\Dbar_k,
\quad \Ubar_i\Dbar_j\Dbar_k,\quad  {\tilde \mu}L_i\Hbar,
\eeq
where ${\tilde \mu}$ is a dimensionful parameter. The indices $i,j,k$ are
generation indices.
For the three singlet solution we obtain the following additional terms
\barr
LL\Ebar:&& {\rm none};\quad \Ubar\Dbar\Dbar:\;\;\udd{3}{1}{3},\,
\udd{2}{2}{3},\\
LQ\Dbar:&& \lqd{1}{1}{2},\,\lqd{1}{2}{1};\,\lqd{3}{1}{3},\,\lqd{3}{3}{1},\,
\lqd{3}{2}{2},
\earr
$LQ\Dbar$ and $\Ubar\Dbar\Dbar$ terms together lead to a dangerous level
of proton decay. We thus exclude the three singlet
solution. Similarly we also exclude the four singlet solutions with $q_1=-1$.
For the ten models of Table~1 \cite{us} we find the
following sets of gauge invariant $R$-parity violating dimension-four terms
\barr
I:&&\lle{1}{3}{3},\,\lqd{1}{3}{3} \quad III:\lle{1}{3}{1}\quad IV:
\lqd{1}{2}{3},\,\lqd{1}{3}{2},\\
V:&&\lqd{1}{1}{3},\,\lqd{1}{3}{1},\quad
VII:\lqd{1}{2}{2},\quad VIII:\lqd{1}{1}{2},\,\lqd{1}{2}{1},\quad
X:L_1\Hbar.\nonum
\earr
We have models with only $LL\Ebar$ type
couplings, others with only $L_i\Hbar$ or $LQ\Dbar$ couplings. We also
have three sets $II,VI,IX$ where \R-parity is conserved. Thus there is no
logical connection between a conserved \R-symmetry and the status of
\R-parity.

The $L_{1,2}\Hbar$ term has a dimensionful coupling $\tilde\mu$
similar to the $\mu$ term of the MSSM. In order to avoid a further
hierarchy problem we require the absence of $L_i\Hbar$ terms and therefore
exclude the models ${X},{X}'$.

Interestingly enough, most of the models predict sizeable $L_{1,2}Q_i\Dbar_j$
interactions. The first set leads to resonant squark production at HERA which
has been investigated in detail in \cite{butter}. This should be observable
with an integrated luminosity of about $100\,pb^{-1}$ for squark masses below
$275\gev$. The second set also lead to observable signals at HERA even for
very small couplings as discussed in \cite{mora}. These models should also
be observable at a hadron collider \cite{rphad}.
We point out that only in model $I$ we have additional terms $L_1HN$. These
conserve \R-parity provided $N$ is interpreted as a right-handed neutrino.
$L_1HN$ is a Dirac neutrino mass and requires a very small Yukawa coupling.
We thus exclude model $I$.
It is interesting to note that eventhough for the Higgs Yukawa couplings the
third generation is dominant this is not necessarily the case for the $R_p$
violating interactions.

It is worth pointing out that models {\bf I, III, IV, V, VII, VIII} are just
of the type postulated in \cite{rpsphal}. In order to maintain GUT-scale
baryogenesis at low-energies despite the sphaleron interactions
and have R-parity violation at a measurable level at least one lepton number
had to be conserved. This is guaranteed by an anomaly-free gauge symmetry in
our models.

Finally we point out that the present work can easily be extended to
include a solution to the mu problem \cite{aliherbi2}. We must drop
the N-field. Then the R-charge of $H_1H_2$ is just 0, so it is
disallowed in the superpotential but allowed in the K\"ahler potential.
The corresponding equations have solutions for 4 extra singlets.
$H_1H_2$ do not couple to Planck scale fields.


\begin{thebibliography}{99}
\bibitem{u1p}{A. Chamseddine and H. Dreiner, ETH-preprint, ETH-TH-95-06;
HEP-PH 9503454.}
\bibitem{us}{A. Chamseddine and H. Dreiner, ETH-preprint, ETH-TH-95-04;
HEP-PH 9504337.}
\bibitem{rinvariance1}{A. Salam and J. Strathdee,
Nucl. Phys. B 87 (1975) 85; P. Fayet Nucl. Phys. B 90 (1975) 104.}
\bibitem{freedman}{D.Z. Freedman, Phys. Rev. D 15 (1977) 1173.}
\bibitem{ferkugo}{S. Ferrara, L. Girardello, T. Kugo and A. van Proeyen,
Nucl. Phys. B (1983) 191.}
\bibitem{gravity}{R. Delbourgo and A. Salam, Phys. lett. B 40 (1972) 381;
T. Eguchi and P. Freund, Phys. Rev. Lett 37 (1976) 1251;
L. Alvarez-Gaume and E. Witten, Nucl. Phys. B 234 (1983) 269.}
\bibitem{gs}{M. B. Green and J. H. Schwarz, Phys. Lett. B 149 (1984) 117.}
\bibitem{alicl2}{A. H. Chamseddine, R. Arnowitt, and P. Nath,
Phys. Rev. Lett. 50 (1983) 232; A. H. Chamseddine, R. Arnowitt  and P. Nath,
Phys. Rev. Lett. 49 (1982) 970.}
\bibitem{cfm}{D. Castano, D. Freedman, and C. Manuel, hep-ph/9507397.}
\bibitem{butter}{J. Butterworth and H. Dreiner, Nucl. Phys. B
397 (1993) 3; Proc. of the $2^{nd}$ HERA Workshop on Physics, 1991.}
\bibitem{mora}{H. Dreiner and P. Morawitz, Nucl. Phys. B 428 (1994) 31.}
\bibitem{rphad}{H. Dreiner and G.G. Ross, Nucl. Phys. B 365 (1991) 597.}
\bibitem{rpsphal}{H. Dreiner and G.G. Ross, Nucl. Phys. B 410 (1993) 188.}
\bibitem{aliherbi2}{A. Chamseddine and H. Dreiner, work in progress.}
\end{thebibliography}
\end{document}